\documentclass[runningheads]{llncs}

\usepackage{graphicx}
\usepackage{amsmath,amssymb}
\usepackage{booktabs}
\usepackage{enumitem}
\usepackage[hidelinks]{hyperref}
\usepackage{microtype}
\usepackage{listings}
\usepackage{orcidlink}
\renewcommand{\orcidID}[1]{\unskip\orcidlink{#1}}

\makeatletter
\renewcommand{\orcidID}[1]{}
\makeatother
\title{Improving Search Suggestions for Alphanumeric Queries}

\title{Improving Search Suggestions for\\Alphanumeric Queries}
\titlerunning{Improving Search Suggestions for Alphanumeric Queries}

\author{
  Samarth Agrawal\inst{1}\orcidID{0009-0008-0197-7717} \and
  Jayanth Yetukuri\inst{1}\orcidID{0000-0002-9204-3889} \and
  Diptesh Kanojia\inst{2}\orcidID{0000-0001-8814-0080} \and
  Qunzhi Zhou\inst{1}\orcidID{0009-0004-8656-6298} \and
  Zhe Wu\inst{1}\orcidID{0009-0003-6109-3386}
}

\authorrunning{S. Agrawal et al.}

\institute{
  eBay Inc., USA\\
  \email{\{samagrawal,jyetukuri,qunzhou,zwu1\}@ebay.com}
  \and
  University of Surrey, Guildford, UK\\
  \email{d.kanojia@surrey.ac.uk}
}

\begin{document}
\maketitle
\begin{abstract}
Alphanumeric identifiers such as manufacturer part numbers (MPNs), SKUs, and model codes are ubiquitous in e-commerce catalogs and search. These identifiers are sparse, non linguistic, and highly sensitive to tokenization and typographical variation, rendering conventional lexical and embedding based retrieval methods ineffective. We propose a training free, character level retrieval framework that encodes each alphanumeric sequence as a fixed length binary vector. This representation enables efficient similarity computation via Hamming distance and supports nearest neighbor retrieval over large identifier corpora. An optional re-ranking stage using edit distance refines precision while preserving latency guarantees.  The method offers a practical and interpretable alternative to learned dense retrieval models, making it suitable for production deployment in search suggestion generation systems. Significant gains in business metrics in the A/B test further prove utility of our approach.

\end{abstract}
\section{Introduction}
Alphanumeric queries such as manufacturer part numbers, stock keeping units, and model codes present a long-standing challenge for search systems. Unlike natural language queries, these identifiers are nonlinguistic, sparse, and highly sensitive to small character-level variations. Conventional lexical retrieval performs poorly because such codes rarely repeat in the corpus, while dense embeddings often blur distinctions between visually similar but semantically distinct strings. Tokenization further complicates the problem because strings like ``S3221QS'' may be segmented inconsistently across models. As a result, even a single character mistake or formatting change can produce irrelevant results, which reduces search relevance and conversion rates.

Existing approaches for improving robustness in retrieval, including character-level embeddings, hybrid neural lexical ranking, and approximate nearest neighbor search in dense space, usually depend on supervised training and significant infrastructure for vector indexing and maintenance. These methods are often too complex for the simple structure of alphanumeric strings and introduce latency.

\begin{figure*}[t]
    \centering
    \includegraphics[width=0.9\linewidth]{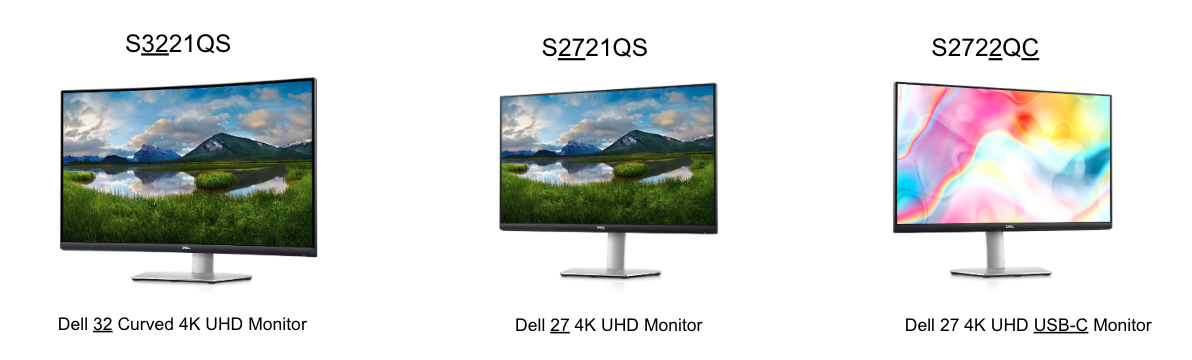}
    \caption{Product model codes vary by a few characters that can correlate with features.}
    \label{fig:monitors}
\end{figure*}

Based on general observations, we hypothesize that product variations (\textit{e.g.}, size, color, interface) typically differ by only a few characters in their model codes. For example in Fig~\ref{fig:monitors}, Dell monitor codes suggest that model numbers vary slightly for different sizes, features, \textit{etc.} implying that nearest neighbors in model code space reflect genuine product family structure.

A training-free nonparametric retrieval framework designed specifically for alphanumeric identifiers is presented. Candidate matches are retrieved by computing Hamming distance between the query vector and an indexed collection of product codes, followed by optional edit distance re-ranking. This procedure enables efficient nearest neighbor retrieval that remains robust to local variations such as typos, version suffixes, or model differences commonly found in product families. When applied as a candidate generator or related query suggester, it delivers fast and accurate results for user inputs that only partially match catalog identifiers.

In summary, this work introduces a lightweight yet effective formulation for alphanumeric retrieval based on binary character encoding and Hamming distance search. The method achieves strong retrieval quality, extremely low latency, and a small memory footprint. It achieves high recall for slightly perturbed queries without requiring any learned parameters.

\paragraph{\textbf{Related Work.}}
Alphanumeric identifiers are fragile under typos and formatting noise. Classic approximate string matching provides tolerance through dynamic programming, bit-vector methods, and practical $q$-gram filtering with metric indexing \cite{navarro2001,myers1999,ukkonen1992}. At catalog scale, non-parametric neighbor search under Hamming distance (e.g., Multi-Index Hashing) and ANN structures (e.g., HNSW) enable sublinear retrieval \cite{mih2012,hnsw}. In e-commerce IR, benchmarks such as ESCI emphasize semantic relevance over identifier recovery \cite{esci}. Recent eBay work studies retrieval for \emph{alphanumeric queries} and broader methods with identifier results \cite{saadany2024product,saadany2024centrality,qian2025near}. For related-search suggestion, prior approaches mine co-clicked queries \cite{Hasan2011QuerySF} or model transitional reformulations with LLMs \cite{yetukuri2025ai}. The proposed method is training-free, identifier-robust, and suitable as a low-latency first-stage generator.

Tokenization-free or byte-level models (\textit{e.g.}, ByT5) address subword brittleness but add training and latency costs \cite{byt5}. Work on query suggestion typically targets auto-completion or ranking rather than identifier-space proximity \cite{qacsurvey}; here, suggestions are derived directly from nearest neighbors in code space.
\section{Methodology}
The proposed methodology comprises two key components: data collection and online inference. The former focuses on building the identifier–query corpus, while the latter addresses real-time retrieval and recommendation.
\begin{figure}[t]
  \centering
      \includegraphics[width=1\linewidth]{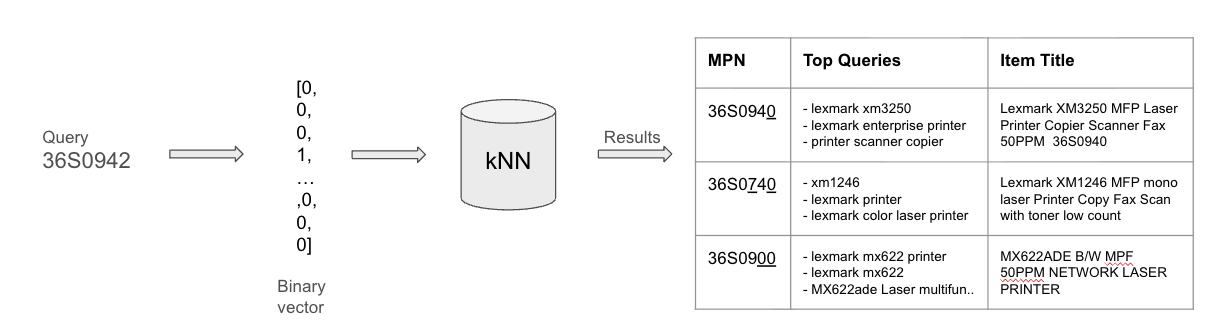}
  \caption{Proposed online inference pipeline for alphanumeric queries.}
  \label{fig:architecture}
\end{figure}

\subsection{Data Collection}
Alphanumeric identifiers such as manufacturer part numbers (MPNs) and model numbers are harvested from current and historical inventory records, normalized, and deduplicated. For each unique code, the \textit{top-selling} item from historical sales is selected. Search log data are then used to associate the \textit{top-3} user queries that resulted in a click or purchase of that item. To reduce identifier collisions, only codes containing $7$ or more characters are retained. Each code is subsequently mapped to a fixed-length 120-bit binary vector by concatenating 6-bit encodings of individual characters (covering \texttt{A–Z}, \texttt{0–9}, and \texttt{-/.}) and padding or truncating to a maximum length of $20$ characters. This representation enables efficient similarity computation via Hamming distance. A Hamming-space approximate nearest-neighbor (ANN) index (\textit{e.g.}, FAISS or Annoy) is constructed, where the binary vector serves as the \textit{key} and the corresponding \textit{value} stores the canonical code along with its associated top-three queries.

\subsection{Online Inference}
Given an input query, the system (i) normalizes and gates it using a regular expression that detects predominantly alphanumeric patterns; if the gate fails, control is passed to the baseline semantic or lexical retrieval stack. Otherwise, the query is (ii) encoded into a fixed-length binary vector, (iii) used to retrieve the top-$k$ neighbors from the Hamming ANN index, (iv) refined by applying a Levenshtein filter on the top-$N$ candidates to tolerate insertions and deletions while preserving close matches, and (v) aggregated and de-duplicated to produce the \emph{top-3} related queries associated with the surviving neighbors. Final suggestions are ranked by a weighted combination of neighbor proximity and historical query frequency or engagement, yielding low-latency, identifier-aware recommendations.

\section{Evaluation}
\begin{figure}[t]
    \centering
    \includegraphics[width=0.9\linewidth]{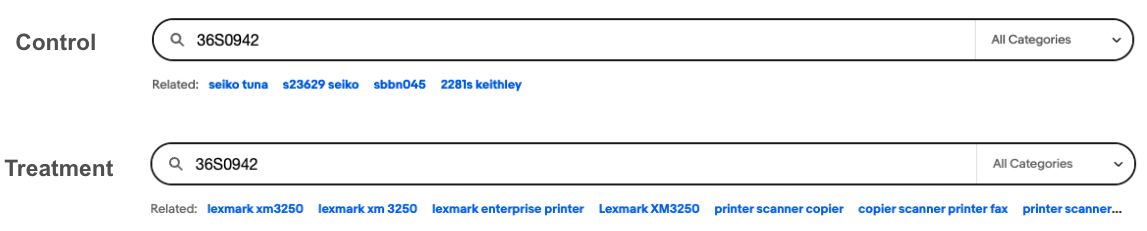}
    \caption{Comparison of search suggestions between control and treatment systems.}
    \label{fig:ab}
\end{figure}

After manually evaluating the quality of generated suggestions, this approach was tested using an online A/B test on production traffic restricted to \emph{alphanumeric queries}. The control rendered the existing related‑search experience; the treatment used our Hamming+edit pipeline to trigger and populate suggestions (Fig: \ref{fig:ab}). As expected, related‑search \emph{coverage} improved by \textbf{+18.8\%}. \emph{Statistically significant} gains of \textbf{+3.35\%} in related‑search CTR and \textbf{+44.4\%} in conversions were observed. These lifts indicate this method surfaces suggestions that users engage with and lead to downstream purchases.

A practical challenge is a `chicken‑and‑egg' effect regarding identifier quality, \textit{i.e.}, many sellers enter arbitrary values in MPN fields because these fields have historically provided limited utility, which in turn makes it harder to effectively exploit MPNs. While our method performs well where codes are reliable, larger gains will require improved normalization/validation at listing time and incentives for accurate MPN entry.

\section{Conclusion}
This work introduced a training-free retrieval method for alphanumeric queries that encodes each identifier as a fixed-length binary vector, retrieves nearest neighbors via Hamming distance, and refines candidates with a lightweight Levenshtein filter. The resulting approximate nearest-neighbor (ANN) index mapping codes to top queries is simple to maintain, low-latency, and complementary to existing semantic and lexical stacks. In online A/B tests on alphanumeric traffic, the approach improved coverage and produced statistically significant gains in click-through rate and conversion. A key limitation lies in the variable quality of seller-provided metadata, which affects query-item associations. Future work will explore using metadata from the ANN index as context for large language models to generate richer offline suggestions without impacting latency.

\section{Speaker bio}
Samarth Agrawal is an MTS2, Software Engineer working on problems
intersecting information retrieval and natural language processing
relevant to the Query Understanding team at
eBay. He is passionate about novel research in
the information retrieval space that can help optimize product
search.

\begin{credits}
\subsubsection{\discintname}
The authors have no competing interests to declare that are relevant to the content of this article.
\end{credits}

\end{document}